\documentclass[prl,aps,nofootinbib,preprintnumbers,superscriptaddress,onecolumn]{revtex4}
\usepackage{textcomp}
\usepackage[marginal]{footmisc}
\usepackage{appendix}
\usepackage{amssymb,amsthm,amsfonts,amsmath,pifont,amscd}
\usepackage{bm}                               
\usepackage{dsfont}
\usepackage{epsfig,eepic,epic,fancybox,multirow}
\usepackage{mathrsfs}
\usepackage{latexsym,rotating}

\usepackage{times} 
\usepackage{mathptmx}

\begin{document}

\title{A comparative study of Gaussian Graphical Model approaches for genomic data}

\author{P.~F.~Stifanelli}\address{Institute of Intelligent Systems for Automation, National Research Council -- CNR, I-70126, Bari, Italy}\address{Physics Department, University of Bari, I-70126, Bari, Italy}

\author{T.~M.~Creanza}\address{Institute of Intelligent Systems for Automation, National Research Council -- CNR, I-70126, Bari, Italy}\address{Department of Emergency and Organ Transplantation -- DETO, University of Bari, I-70124, Bari, Italy}

\author{R.~Anglani}\address{Institute of Intelligent Systems for Automation, National Research Council -- CNR, I-70126, Bari, Italy}

\author{V.~C.~Liuzzi}\address{Institute of Intelligent Systems for Automation, National Research Council -- CNR, I-70126, Bari, Italy}

\author{S.~Mukherjee}\address{Institute for Genome Sciences and Policy, Center for Interdisciplinary Engineering, Medicine and Applied Sciences, Duke University, 101 Science Drive, Durham, NC 27708, USA }
\author{N.~Ancona}\address{Institute of Intelligent Systems for Automation, National Research Council -- CNR, I-70126, Bari, Italy}\email{nicola.ancona@ba.issia.cnr.it}

\begin{abstract}
The inference of networks of dependencies by Gaussian Graphical
models on high-throughput data is an open issue in modern molecular
biology. In this paper we provide a comparative study of three
methods to obtain small sample and high dimension estimates of
partial correlation coefficients: the Moore-Penrose pseudoinverse
(PINV), residual correlation (RCM) and covariance-regularized method
$(\ell_{2C})$. We first compare them on
simulated datasets and we find that PINV is less stable in
terms of AUC performance when the number of variables changes. The
two regularized methods have comparable performances but $\ell_{2C}$
is much faster than RCM. Finally, we present the results of an
application of $\ell_{2C}$ for the inference of a gene network for
isoprenoid biosynthesis pathways in {\it Arabidopsis thaliana}.
\end{abstract}
\preprint{BA-TH/643-11}

\maketitle

\section{Introduction}
One of the aims of systems biology is to provide quantitative
models for the study of complex interaction patterns among genes
and their products that are the result
 of many biological processes in the cell, such as biochemical interactions and regulatory activities.
 In this framework, graphical models \cite{Lauritzen_book} have been
  exploited as useful stochastic tools to investigate and describe the conditional independence structure between random variables.
   In particular, the Graphical Gaussian Models (GGM) use the partial correlation estimates as a measure of conditional
   independence between any two variables \cite{Dempster1972}. Unfortunately, the application of GGMs classical theory is still a hard task. The genomic data are tipically characterized by a huge number of genes $p$ with respect to the small number of available samples $n$. This makes unreliable the application of the classical GGMs theory to the small sample setting case. In recent years, several methods have been proposed to overcome this
problem by reducing the numbers of genes or gene lists in order to
reach the $n>p$ regime \cite{Toh2002}. Other solutions have been
also proposed \cite{Wille2004,Castelo2006,Gilbert2009} to
circumvent the problem of computing full partial correlation
coefficients by using only zero and first order coefficients.
However, these approaches do not take into account all multigene
effects on each pair of variables. A more sophisticated way to
adapt GGMs to the $n<p$ case is to find regularized estimates for
the covariance matrix \cite{Yuan2007,Friedman2008,Witten2009} and
its inverse. Once regularized estimates of partial correlation are
available, heuristic searches can be used to find an optimal
graphical model. A fundamental assumption to perform these
quantitative methods is the sparsity of biological networks: only
a few edges are supposed to be present in the gene regulatory
networks, so that reliable estimates of the graphical model can be
inferred also in small sample case \cite{Castelo2006}. A
regularized GGM method based on a Stein-type shrinkage has been
applied to genomic data \cite{Dobra2004} and the network selection
has been based on false discovery rate multiple testing. In
Ref.~\cite{Schaffer2005} the same procedure to select the network
has been adopted, with a Moore-Penrose pseudoinverse method to
obtain the concentration matrix. Finally, the authors in
Ref.~\cite{Meinshausen2006} have suggested an attractive and
simple approach based on lasso-type regression to select among the
partial correlations the nonzero values, paving the way to a
number of analysis and novel algorithms based on lasso $\ell_{1}$
regularizations
\cite{Yuan2007,Friedman2008,Witten2009,Friedman2010}. In this
work, we focus on regularized methods for the estimation of the
concentration matrix in an undirected GGM. In particular, we
present a comparative study of three methods in terms of AUC (area
under the Receiving Operative Characteristic curve) and timing
performances. One is based on Moore--Penrose pseudoinverse (PINV),
the other two provide an estimate of the partial correlation
coefficients, based on Regularized Least Square regression (RCM)
and a covariance-regularized method with a $\ell_{2}$ penalty in
the log-likelihood function $(\ell_{2C})$. Finally, we apply the
$\ell_{2C}$ method to infer a gene network for the isoprenoid
biosynthesis pathways in {\it A. thaliana}. This network
structural analysis allows to enlight some expected pathway
properties. In particular, we find a negative partial correlation
coefficient between the two hubs in the two isoprenoid pathways.
This suggests a different response of the pathways to the several
tested experimental conditions and, together with the high
connectivity of the two hubs, provides an evidence of cross-talk
between genes in the plastidial and the cytosolic pathways.

\section{Gaussian networks from microarray data}
Let ${\bf X}=(X_1, \dots, X_p)\in \mathbb{R}^{p}$ be a random
vector distributed according a multivariate normal distribution
$\mathcal{N}({\bm \mu},{\bm \Sigma})$. The interaction structure between these
variables can be described by means of a graph $G=(V,E)$, where
$V$ is the vertex set and $E$ is the edge set. If vertices of $V$
are identified with the random variables $X_1, \dots, X_p$, then
the edges of $E$ can represent the conditional dependence between
the vertices. In other words, the absence of an edge between the
$i-$th and $j-$th vertex means a conditional independence between
the associated variables $X_i$ and $X_j$. In this study, we shall
consider only undirected Gaussian graphs $G$ with {\it pairwise Markov
property}, such that for all $(i,j)\notin E$ one has
\begin{align}\label{eq:markov}
 X_i &\perp\!\!\!\!\perp  X_j \;|\; X_{V \backslash\{ i,j \} } &
i,j&=1,\dots,p\,,
\end{align}
i.e. $X_i$ and $X_{j}$ are conditionally independent being fixed
all other variables $X_{V \backslash\{i,j\}}$. Since ${\bf X}$
follows a $p-$variate normal distribution, the condition
\eqref{eq:markov} turns out to be $\rho_{ij\cdot
V\backslash\{i,j\}}=0$, where $\rho_{ij\cdot V\backslash\{i,j\}}$
is the partial correlation coefficient between the $i-$th and
$j-$th variable, being fixed all other variables. It has been
shown \cite{Lauritzen_book} that partial correlation matrix
elements are related to the {\it precision matrix} (or inverse
covariance matrix) $\bm \Theta = \bm \Sigma^{-1}$, as:
\begin{align}\label{eq:par_corr_inv_cov}
\rho_{ij\cdot V\backslash\{i,j\}} =
-\frac{\theta_{ij}}{\sqrt{\theta_{ii}\theta_{jj}}}\;\;\;\;\;i\neq
j\,,
\end{align}
where $\theta_{ij}$ are elements of $\bm \Omega$. In general, when
the number of observations $n$ is greater than the number of
variables $p$, it is straightforward to evaluate $\theta_{ij}$ in
Eq.~\eqref{eq:par_corr_inv_cov} by inverting the sample covariance
matrix. Unfortunately, a typical genomic dataset is characterized
by $n < p$, so that the sample covariance matrix becomes not
invertible \cite{Dikstra1970}. For this reason, in order to
estimate the partial correlation matrix one needs alternative
methods to overcome the problem, like regularization methods,
ridge regression or pseudoinverse.

\subsection{Partial correlation matrix estimation}\label{subs:methods}
In order to describe the three methods that we shall investigate, let us consider the $n\times p$ matrix ${\bf X}=({\bf  X}_{1},{\bf  X}_{2},\dots,{\bf  X}_{p})$, where each $\{{\bf X}_{i}\}\in \mathbb{R}^{n}$,
with $n<p$. Let us indicate $\bf S$ as the estimate of
the covariance matrix ${\bm \Sigma}$ and $\hat{\bm \Theta}$ as the
estimate of inverse covariance matrix ${\bm \Sigma}^{-1}$.

\subsubsection{Pseudoinverse method $(\rm PINV)$}
The precision matrix $\hat{\bm\Theta}$ can be obtained as
pseudoinverse of $\bf S$, by using the Singular Value
Decomposition (SVD). Indeed, a singular value decomposition of a
$m\times q$ matrix $M$, is $M=U\Lambda V^{\ast}\,,$ where $U$ is
a $m\times m$ unitary matrix, $\Lambda$ is $m\times q$ diagonal
matrix with nonnegative real numbers on the diagonal and
$V^{\ast}$ is a $q\times q$ unitary matrix (transpose conjugate of
$V$). Then, the pseudoinverse of $M$ is
$M^{+}=V\Lambda^{+}U^{\ast}$, where $\Lambda^+$ is obtained by
replacing each diagonal element with its reciprocal and then
transposing the matrix.

\subsubsection{Covariance-regularized method $(\ell_{2C})$}
Let us consider a log likelihood function with a $\ell_2$ penalization
\cite{Witten2009}:
\begin{align}\label{eq:log-likelihood}
L({\bm \Theta})=\log\det{\bm \Theta}-{\rm Tr}({\bf S}{\bm \Theta})
-\lambda\|{\bm \Theta}\|_F^2\,,
\end{align}
with $\lambda>0$ and $\|{\bm \Theta}\|^2_F={\rm tr}({\bm
\Theta^{\top}{\bm \Theta}})$. The maximization of
Eq.~\eqref{eq:log-likelihood} with respect to $\bm \Theta$
is equivalent to solve the following equation
\begin{equation}\label{eq:eigenvalue_problem}
\hat{\bm \Theta}^{-1}-2\lambda\hat{\bm \Theta}={\bf S}\,.
\end{equation}
Consequently, the problem turns out to be an eigenvalue problem,
therefore the eigenvalues $\theta_i$ of $\hat{\bm \Theta}$ can be
evaluated as function of the eigenvalues $s_i$ of ${\bf S}$:
\begin{equation}
    \theta_i^{\pm}=-\frac{s_i}{4\lambda}\pm
    \frac{\sqrt{s_i^2+8\lambda}}{4\lambda}\,.
\end{equation}
Since ${\bm \Theta}$ must be positive definite, the correct
value of $\theta_i$ is $\theta^+_{i}$ then, for the spectral theorem
the precision matrix $\hat{\bm \Theta}$ is given by
\begin{equation}
    \hat{\bm \Theta}=\sum_{i=1}^{\ell}\theta^+_i{\bf u}_i{\bf u}_i^\top\,.
\end{equation}
Finally, in order to estimate the parameter $\lambda$ that maximizes
the penalized log-likelihood function in
Eq.~\eqref{eq:log-likelihood}, we carry out 20 random splits of the
data set in training and validation sets and then we evaluate the
log-likelihood over the validation set.

\subsubsection{Residual correlation method $(\rm RCM)$}
We consider a regression model for the
variables ${\bf X}_i$ and ${\bf X}_j$ as
\begin{align}\label{eq:regr_model}
{\bf X}_{i}&=\langle {\bm \beta}_{(i)}, {\bf X}_{\backslash i
\backslash j}\rangle + b_i & {\bf X}_{j}&=\langle {\bm
\beta}_{(j)}, {\bf X}_{\backslash i \backslash j}\rangle + b_j
\end{align}
where $\{{\bm \beta}_{(i)}\}$ is the regression coefficient vector
in $p-2$ dimensions referred to the $i-$th gene; ${\bf X}_i$ is
the $i-$th column of the matrix ${\bf X}$ and ${\bf X}_{\backslash
i\backslash j}$ is ${\bf X}$ without the $i-$th and $j-$th
columns. The Regularized Least Square (RLS) \cite{RLS} method evaluates the
regression models \eqref{eq:regr_model} by solving
\begin{align}
\min_{\beta\in\mathbb{R}^{p-2}}\frac{1}{n}\|{\bf X}_i - {\bm
\beta}_{(i)}{\bf X}_{\backslash i \backslash j}\|_2^2+\lambda\|\bm
\beta_{(i)}\|^2_2\,.
\end{align}
Now, if $\tilde {\bf X}_i$ and $\tilde {\bf X}_j$ are the RLS
estimates of ${\bf X}_i$ and ${\bf X}_j$, one can evaluate the
residual vectors $ {\bf r}_{i}=\tilde{\bf X}_i-{\bf X}_i$ and
${\bf r}_{j}=\tilde{\bf X}_j-{\bf X}_j$. This allows to evaluate
the partial correlation coefficients $\rho_{ij|p-2}$ between the
$i-$th and $j-$th variable being fixed all other $p-2$ variables
as the Pearson correlation $r_{r_{i}r_{j}}$ between the residuals,
i.e.
\begin{equation}\label{eq:pearson_residual}
\rho_{ij|p-2}=r_{r_{i}r_{j}}=\frac{\rm cov({\bf r}_i,{\bf
r}_j)}{\sqrt{\rm var({\bf r}_i)\cdot \rm var({\bf r}_i)}}\,.
\end{equation}
Finally, the $\lambda>0$ parameter has been chosen by minimizing the
Leave-One-Out cross validation errors.

\section{Comparative study of accuracy}

\subsection{Data generation}\label{subs:data_generation}
Datasets with different numbers of variables and observations have
been used in order to investigate the performances of the methods,
i.e. $p=\{50,200,400\}$ and $n=\{20,200,500\}$. Each dataset $\bf
X$ has been generated from a multivariate gaussian distribution
with zero mean and covariance ${\bm \Sigma}_{\rm th} = {\bm
\Theta}_{\rm th}^{-1}$. The structure of the precision matrix
${\bm \Theta}_{\rm th}$ presents the following patterns
\cite{Friedman2010}: {\it random}, {\it hubs} and {\it cliques}
and it has approximately $p$ non vanishing entries out of the
$p(p-1)/2$ off-diagonal elements, except for clique configuration
where the entries are approximately $2p$.

In the {\it random} pattern, the off-diagonal terms of ${\bm
\Theta}_{\rm th}$ are set randomly to a fixed value $\theta \neq
0$. In the {\it hubs} configuration, we partition the columns into
disjoint groups $G_k$, where index $k$ indicates the $k-$th column
chosen as ``central'' in each group. Then the off-diagonal terms
are set $\theta_{ik}=\theta$ if $i\in G_k$, otherwise
$\theta_{ik}=0$. In the {\it cliques} pattern, the precision
matrix is partitioned as done in {\it hubs} and the off-diagonal
terms $\theta_{ij}$ are set to $\theta$ if $i,j\in G_k$, with
$i\neq j$. The positive definiteness for each configuration, is
guaranteed by the diagonal entries which are selected in order to
keep ${\bm \Theta}_{\rm th}$ diagonally dominant.

\begin{table}\begin{center}\noindent
\begin{tabular}{|ccc|}
\hline
      &&       \\
      &&   n   \\
 \hline\hline
  $r$ &&  {500}\\
  $h$ &&  {500}\\
  $c$ &&  {500}\\
 \hline
  $r$ &&  {200}\\
  $h$ &&  {200}\\
  $c$ &&  {200}\\
 \hline
  $r$ &&   {20}\\
  $h$ &&   {20}\\
  $c$ &&   {20}\\
 \hline
 \hline
\end{tabular}~\begin{tabular}{|ccccccc|}
\hline
 &&      && $\ell_{2C}$  &&\\
\hline
 &&  AUC &&  AUC std     && T (s) \\
\hline\hline
 &&   0.998 && 0.0001 && 38.86 \\
 &&   1.000 && 0.0000 && 83.74 \\
 &&   0.995 && 0.0002 && 84.95 \\
\hline
 &&   0.976 && 0.0003 && 38.44 \\
 &&   1.000 && 0.0000 && 81.13 \\
 &&   0.936 && 0.0008 && 82.02 \\
\hline
 &&   0.808 && 0.0011 && 39.03 \\
 &&   0.999 && 0.0001 && 82.03 \\
 &&   0.668 && 0.0014 && 82.13 \\
\hline\hline
\end{tabular}~\begin{tabular}{|ccccccc|}
  \hline && && PINV  &&\\\hline
 && AUC &&AUC std &&  T (s) \\
 \hline\hline
 && 0.987 && 0.0006 && 0.161 \\
 && 0.999 && 0.0000 && 0.164 \\
 && 0.963 && 0.0014 && 0.164 \\
 \hline
 && 0.581 && 0.0161 && 0.111 \\
 && 0.806 && 0.0150 && 0.115 \\
 && 0.587 && 0.0049 && 0.121 \\
 \hline
 && 0.929 && 0.0018 && 0.093 \\
 && 1.000 && 0.0000 && 0.091 \\
 && 0.659 && 0.0014 && 0.091 \\
 \hline\hline
\end{tabular}~\begin{tabular}{|ccccccc|}
  \hline && && RCM  &&\\\hline
 && AUC &&AUC std &&T (s) \\
  \hline\hline
 && 0.999 && 0.0001 && 8343 \\
 && 1.000 && 0.0000   && 6468 \\
 && 0.996 && 0.0002 && 6449 \\
 \hline
 && 0.984 && 0.0006 && 3566 \\
 && 0.999 && 0.0001 && 3555 \\
 && 0.923 && 0.0009 && 3747 \\
 \hline
 && 0.924 && 0.0017 && 105 \\
 && 0.999 && 0.0000      && 106 \\
 && 0.659 && 0.0014 && 108 \\
 \hline\hline
\end{tabular}
\end{center}
\caption{AUC, AUC standard error and timing performances for
$p=400$. {\it Left part}: $\ell_{2C}$ method. {\it Center part}:
PINV. {\it Right part}: RCM. Indices $r$, $h$ and $c$ stand for
random, hubs and clique pattern, respectively. }\label{tab:p400}
\end{table}

\begin{table}\begin{center}
\begin{tabular}{|ccc|}
      \hline &&       \\
      &&   n   \\
 \hline\hline
  $r$ &&  {500}\\
  $h$ &&  {500}\\
  $c$ &&  {500}\\
 \hline
  $r$ &&  {200}\\
  $h$ &&  {200}\\
  $c$ &&  {200}\\
 \hline
  $r$ &&   {20}\\
  $h$ &&   {20}\\
  $c$ &&   {20}\\
 \hline
 \hline
\end{tabular}~\begin{tabular}{|ccccccc|}
\hline
 &&         && $\ell_{2C}$  &&\\
\hline
 && AUC     && AUC std && T (s) \\
\hline\hline
 &&   0.999 && 0.0001 && 5.807 \\
 &&   1.000 && 0.0000        && 10.655 \\
 &&   0.996 && 0.0002 && 10.821 \\
\hline
 &&   0.986 && 0.0003 && 5.592 \\
 &&   1.000        && 0.0000        && 10.425 \\
 &&   0.944 && 0.0010 && 10.529 \\
\hline
 &&   0.784 && 0.0016 && 6.150 \\
 &&   0.999 && 0.0001 && 10.574 \\
 &&   0.669 && 0.0016 && 10.545 \\
\hline\hline
\end{tabular}~\begin{tabular}{|ccccccc|}
  \hline && && PINV  &&\\ \hline
 &&  AUC &&AUC std &&  T (s) \\
  \hline\hline
 && 0.999 && 0.0001 && 0.0377 \\
 && 1.000        && 0.0000   && 0.0376 \\
 && 0.999 && 0.0001 && 0.0439 \\
 \hline
 && 0.703 && 0.0067 && 0.0310 \\
 && 0.748 && 0.0124 && 0.0309\\
 && 0.612 && 0.0064 && 0.0336 \\
 \hline
 && 0.880 && 0.0048 && 0.0187 \\
 && 0.999 && 0.0002 && 0.0182 \\
 && 0.649 && 0.0017 && 0.0189 \\
 \hline\hline
\end{tabular}~\begin{tabular}{|ccccccc|}
  \hline && && RCM  &&\\ \hline
 && AUC && AUC std && T (s) \\
  \hline\hline
 && 0.999 && 0.0001 && 807 \\
 && 1.000        && 0.0000     && 450 \\
 && 0.999 && 0.0000 && 436 \\
 \hline
 && 0.990 && 0.0007 && 861 \\
 && 0.999 && 0.0003 && 856 \\
 && 0.950 && 0.0008 && 1028 \\
 \hline
 && 0.871 && 0.0046 && 24.5 \\
 && 0.999 && 0.0001 && 27.9 \\
 && 0.654 && 0.0017 && 25.3 \\
 \hline\hline
\end{tabular}
\end{center}
\caption{AUC, AUC standard error and timing performances for
$p=200$. {\it Left part}: $\ell_{2C}$ method. {\it Center part}:
PINV. {\it Right part}: RCM. Indices $r$, $h$ and $c$ stand for
random, hubs and clique pattern, respectively. }\label{tab:p200}
\end{table}

\subsection{Performances} In order to compare the performances of
the three methods, we have used this procedure: (I) For each data
generation pattern, draw a random dataset ${\bf X}$ from
$\mathcal{N}({\bf 0},{\bm \Sigma}_{\rm th})$; (II) Evaluate ${\bf
S}$ and $\bm \Theta_{\exp}$ in the case of PINV and $\ell_{2C}$,
hence find $\rho_{\rm exp}$ from Eq.~\eqref{eq:par_corr_inv_cov};
in the case of RCM use Eq.~\eqref{eq:pearson_residual} for the
evaluation of $\rho_{\rm exp}$; (III) For each method, evaluate
the AUC performance, as follows. Since the edges in our simulated
dataset have the same strength and we know the label edge and non
edge for each element, the elements of $\rho_{\rm exp}$ can be
divided in two sets: $\rho_{\rm exp}$ for the edge elements and
$\rho_{\rm exp}$ for the non edge ones. The AUC measures the
performances of the three methods in terms of accuracy of
classification of edge and non edges by using the relative
$\rho_{\rm exp}$  values.

\section{Results}
In Tables \ref{tab:p400}, \ref{tab:p200} and \ref{tab:p50} we
present the AUC, AUC standard error and timing (in seconds)
performances for $p=\{400,200,50\}$, respectively. Each table is
divided in three columns related to the analyzed methods. Indices
$r$, $h$, and $c$ refer to the three data generation methods:
random, hubs, and clique. The results shown are averaged over 20
trials for $n=\{500,200,20\}$.

As expected, when $n>p$ all methods provide the same efficiency
with an AUC virtually equal to 1. In fact, in this case the use of
regularization methods should be not required. When $p>n$, we find
that PINV presents some instability in AUC outcomes, mainly in
those region when $p\approx n$. This can be due to a ``resonance
effect'', as explained in Refs.~\cite{Schaffer2005,Raudys1998}.
Instead, RCM and $\ell_{2C}$ show high value of AUC in all
settings and have similar performances, almost indipendently of
the range of $p$ and $n$. Note that, only in the random
configuration, when $n=20$ and $p=\{200,400\}$, RCM shows AUC
values 10\% larger than $\ell_{2C}$ ones. On the other hand, the
timing comparison highlights that $\ell_{2C}$ is much faster than
the RLS-based method.

\section{Application to biological pathways}
Isoprenoids play various important roles in plants, functioning as
membrane components, photosynthetic pigments, hormones and plant
defence compounds. They are synthesized through condensation of
the five-carbon intermediates isopentenyl diphosphate (IPP) and
dimethylallyl diphosphate (DMAPP). In higher plants, IPP and DMAPP
are synthesized through two different routes that take place in
two distinct cellular compartments. The cytosolic pathway, also
called MVA (mevalonate) pathway, provides the precursors for
sterols, ubiquinone and sesquiterpenes  \cite{Disch1998}. An
alternative pathway, called MEP/DOXP (2-C-methyl-D-erythritol
4-phosphate / 1-deoxy-D-xylulose 5-phosphate), is located in the
chloroplast and is used for the synthesis of isoprene,
carotenoids, abscisic acid, chlorophylls and plastoquinone
\cite{Lichtenthaler1997}. Although this subcellular
compartmentation allows both pathways to operate independently,
there are several evidences that they can interact in some
conditions \cite{Laule2003}. Inhibition of the MVA pathway in {\it
A. thaliana} leads to an increase of carotenoids and chlorophylls
levels, demonstrating that its decreased functioning can be
partially compensated for by the MEP/DOXP pathway. Inversely,
inhibition of the MEP/DOXP pathway in seedlings causes the
reduction of levels in carotenoids and chlorophylls, indicating a
unidirectional transport of isoprenoid intermediates from the
chloroplast to the cytosol. In order to investigate whether the
transcriptional regulation is at the basis of the crosstalk
between the cytosolic and the plastidial pathways, Laule et al.
\cite{Laule2003} have studied this interaction by identifying the
genes with expression levels changed as a response to the
inhibition. They have shown that the inhibitor mediated changes in
metabolite levels are not reflected in changes in gene expression
levels, suggesting that alterations in the flux through the two
isoprenoid pathways are not transcriptionally regulated. In order
to clarify the interaction between both pathways at the
transcriptional level, Wille et al. \cite{Wille2004} have explored
the structural relationship between genes on the basis of their
expression levels under different experimental conditions. This
study aims to infer the regulatory network of the genes in the
isoprenoid pathways by incorporating the expression levels of 795
genes from other 56 metabolic pathways. Moving beyond the one-gene
approach,
the authors have found various connections between genes in the
two different pathways, suggesting the existence of a crosstalk at
the transcriptional level.

\begin{table}\begin{center}
\begin{tabular}{|ccc|}
     \hline &&       \\
      &&   n   \\
 \hline\hline
  $r$ &&  {500}\\
  $h$ &&  {500}\\
  $c$ &&  {500}\\
 \hline
  $r$ &&  {200}\\
  $h$ &&  {200}\\
  $c$ &&  {200}\\
 \hline
  $r$ &&   {20}\\
  $h$ &&   {20}\\
  $c$ &&   {20}\\
 \hline
 \hline
\end{tabular}~\begin{tabular}{|ccccccc|}
\hline
 &&     && $\ell_{2C}$ &&\\
\hline
 && AUC &&  AUC std    &&  T (s) \\
\hline\hline
 &&   0.999 && 0.0000 && 0.4401 \\
 &&   1.000     && 0.0000     && 0.4506 \\
 &&   0.999 && 0.0000 && 0.4184 \\
\hline
 &&   0.996 && 0.0004 && 0.4206 \\
 &&   1.000        && 0.0000        && 0.4266 \\
 &&   0.976 && 0.0023 && 0.3971 \\
\hline
 &&   0.821 && 0.0047 && 0.4106 \\
 &&   1.000 && 0.0000 && 0.4174 \\
 &&   0.675 && 0.0052 && 0.3776 \\
\hline\hline
\end{tabular}~\begin{tabular}{|ccccccc|}
  \hline&& && PINV  &&\\ \hline
 &&  AUC && AUC std &&  T (s) \\
  \hline\hline
 && 1.000 && 0.0000 && 0.0152 \\
 && 1.000 && 0.0000 && 0.0061 \\
 && 1.000 && 0.0000 && 0.0065 \\
 \hline
 && 0.997 && 0.0004 && 0.0038 \\
 && 1.000 && 0.0000 && 0.0030\\
 && 0.985 && 0.0009 && 0.0036 \\
 \hline
 && 0.654 && 0.0097 && 0.0024 \\
 && 0.542 && 0.0076 && 0.0019 \\
 && 0.574 && 0.0076 && 0.0022 \\
 \hline\hline
\end{tabular}~\begin{tabular}{|ccccccc|}
  \hline&& && RCM  &&\\ \hline
 && AUC && AUC std && T (s) \\
  \hline\hline
 && 1.000 && 0.0000 && 2.76 \\
 && 1.000 && 0.0000        && 4.19 \\
 && 1.000 && 0.0000 && 3.45 \\
 \hline
 && 0.998 && 0.0004 && 1.92 \\
 && 1.000 && 0.0000 && 2.26 \\
 && 0.978 && 0.0011 && 2.10 \\
 \hline
 && 0.815 && 0.0066 && 1.56 \\
 && 0.866 && 0.0081 && 1.43 \\
 && 0.666 && 0.0057 && 1.48 \\
 \hline\hline
\end{tabular}
\end{center}
\caption{AUC, AUC standard error and timing performances for
$p=50$. {\it Left part}: $\ell_{2C}$ method. {\it Center part}:
PINV. {\it Right part}: RCM. Indices $r$, $h$ and $c$ stand for
random, hubs and clique pattern, respectively. }\label{tab:p50}
\end{table}

\subsection{Results from the covariance-regularized method for A. thaliana isoprenoid pathways}
\begin{figure}[t]
\begin{center}
\includegraphics[height=8cm]{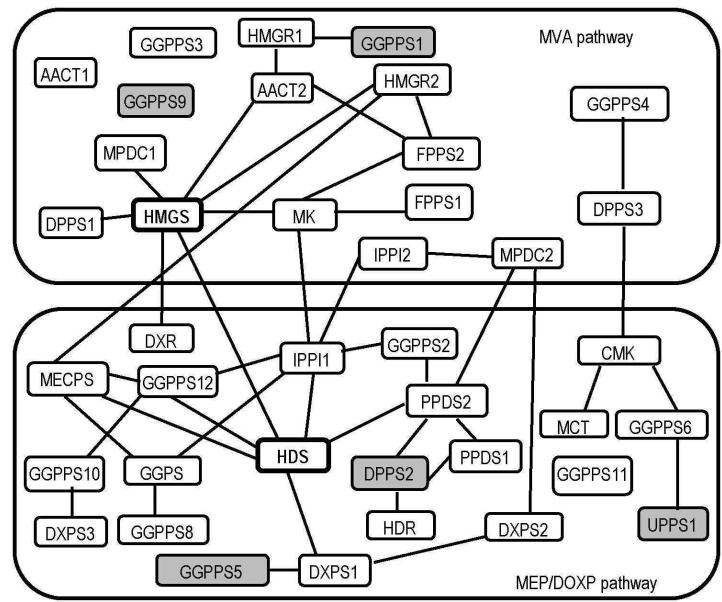}
\caption{Biological network of the isoprenoid pathways inferred by
using PLLM. {\it Upper part}: Genes of MVA pathway. {\it Lower
part}: Genes of MEP/DOXP pathway. Grey boxes refer to
mithochondrial genes; HMGS and HDS represent the hubs of the two
modules.} \label{fig:arabidopsis}
\end{center}
\end{figure}

We apply the $\ell_{2C}$ method to the publicly available data set
from Ref.~\cite{Wille2004}. The selection of the graph is
performed by computing the 95\% bootstrap confidence interval of
the statistics and the absence of an edge occurs when the zero is
included in this interval. The data consist of expression
measurements for 39 genes in the isoprenoid pathways and 795 in
other 56 pathways assayed on 118 Affymetrix GeneChip microarrays.
We are interested in the construction of a gene network in the two
isoprenoid pathways in order to detect the effects of genes in the
other pathways. In Fig.~\ref{fig:arabidopsis} we reproduce the
inferred network with 44 edges. For each pathway we find a module
with strongly interconnected and positively correlated genes. This
suggests the reliability of our method since genes within the same
pathway are potentially jointly regulated \cite{Ihmels2004}.
Furthermore, we identify two strong candidate genes for the
cross-talk between the pathways: HMGS and HDS. HMGS represents the
hub of the cytosolic module, because it is positively correlated
to five genes of the same pathway: DPPS1, MDPC1, AACT2, HMGR2 and
MK. It encodes a protein with hydroxymethylglutaryl-CoA synthase
activity that catalyzes the second step of the MVA pathway. HDS
represents the hub of the plastidial module, because it is
positively correlated to five genes of the same pathway: DXPS1,
MECPS, GGPPS12, IPPI1 and PPDS2. It encodes a
chloroplast-localized hydroxy-2-methyl-2-(E)-butenyl 4-diphosphate
synthase and catalyzes the penultimate step of the biosynthesis of
IPP and DMAPP via the MEP/DOXP pathway. The negative correlation
between HMGS and HDS means that they respond differently to the
several tested experimental conditions. This, together with the
high connectivity of the two hubs, provides an evidence of
cross-talk between genes in the plastidial and the cytosolic
pathways. Other negative correlations between the two pathways are
represented by the edges HMGR2--MECPS, MPDC2--PPDS2 and
MPDC2--DXPS2. Interestingly, the plastidial gene IPPI1 is found to
be positively correlated to the module of connected genes in the
MVA pathway (IPPI1--MK, IPP1--IPPI2). This evidence confirms the
results of Ref.~\cite{Gilbert2009} where they guess that the
enzyme IPPI1 controls the steady-state levels of IPP and DMAPP in
the plastid, when a high level of transfer of intermediates
between plastid and cytosol takes place. Moreover, our study shows
three candidate mitochondrial genes for the cross-talk (DPPS2,
GGPPS5 and UPPS1) which are in the plastidial module. Finally, it
is interesting to note that the method used in
Ref.~\cite{Wille2004} includes more cross-links between the two
pathways with respect to the $\ell_{2C}$ method. Although from the
literature it is known the existence of an interaction between the
two pathways, we believe that this cross-link should not be so
strong, as genes of the two pathways belong to two different cell
compartments. A possible explanation of such a difference is that
Wille {\it et al.} construct a network based on the first-order
conditional dependence that may not capture multi-gene effects on
a given pair of genes.

\section{Conclusions} In this paper, we present a comparative
study of three different methods to infer networks of dependencies
by estimates of partial correlation coefficients in the typical
situation when $n<p$. In particular, we consider the Moore-Penrose
pseudoinverse method (PINV), the residual correlation method (RCM)
and a covariance-regularized method $(\ell_{2C})$. Firstly, we
evaluate AUCs and timing performances on simulated datasets and we
find that PINV presents some instability in AUC outcomes
associated to the variable number variations. On the other hand,
the two regularized methods show comparable performances with a
sensible gain of time elapsing of $\ell_{2C}$ with respect to RCM.
Finally, we present the results of an application of $\ell_{2C}$
for the inference of a gene network for isoprenoid pathways in
 {\it A. thaliana}. We find a negative partial correlation
coefficient between HMGS and HDS, that are the two hubs in the two
isoprenoid pathways. This means that they respond differently to
the several tested experimental conditions and, together with the
high connectivity of the two hubs, provides an evidence of
cross-talk between genes in the plastidial and the cytosolic
pathways. This evidence did not result from studies at level of
single gene. Moreover, studies that infer this network by using
only low-order partial correlation coefficients find more
interactions between the two pathways with respect to the
$\ell_{2C}$ method. A reduced number of edges between the two
pathways is plausible considering the different cell
compartmentalization of the two isoprenoid biosynthesis pathways.
\\

\begin{acknowledgements} This work was supported by grants from
Regione Puglia PO FESR 2007--2013 Progetto BISIMANE (Cod.~n.~44).
\end{acknowledgements}

\end{document}